\documentclass{emulateapj}
\usepackage{natbib}
\usepackage{amsmath}
\usepackage{float}
\usepackage{subfigure}
\usepackage{xcolor}
\usepackage{mathrsfs}
\usepackage{hyperref}
\usepackage{bm}
\usepackage{graphicx}
\usepackage{epstopdf}
\usepackage {color}

\def\be{\begin{equation}}
\def\ee{\end{equation}}
\catcode`\@=11

\shorttitle{}
\shortauthors{}

\begin{document}
\bibliographystyle{apj}
\title
{Pattern of lensed chirp gravitational wave signal and its implication on the mass and position of lens}

\author {Dongze Sun\altaffilmark{1}, Xilong Fan\altaffilmark{2}}
 \altaffiltext{1}{Hongyi Honor School, Wuhan University}
 \altaffiltext{2}{School of Physics and Technology, Wuhan University}

\begin{abstract}
We prose a new measurement strategy of the estimation   the lens mass,  as well as the actual amplification,  of the lens through the modulation patten of  lensed gravitational wave signals alone.     This can be done by  measuring the frequency and  amplitude ratio  of  peaks and valleys of the  modulation pattern of CBC signals , known as  ``beat",   which is a time domain phenomenon of strong lens effect.
 \end{abstract}

\section{INTRODUCTION}
The detection of gravitational waves (GWs) by LIGO/Virgo collaborations \cite{B. P. Abbott et al.}  open the new era of the GW astronomy and the multi-messenger astronomy. If gravitational waves pass through a massive body in the process of propagation, they can be gravitationally lensed \cite{P. Schneider}.
The effects of lensed the tensor wave nature of GW (e.g. rotation  of the polarization plane of GW \cite{hou19a}  ) is  negligible  for most lensed systerms.  Only the amplification effect (e.g. \cite{Takahashi}) of the lens is the  investigated in most literature.

 In the context of 3rd generation interferometric ground detectors - the Eistein Telescope in particular - quite recent series of papers
 explored the perspectives of observing gravitationally lensed coalescing double compact objects   by
the ET for the  SIS lens model \citep{Piorkowska13, Biesiada14, Ding15} and more realistic lens model \cite{Mao}, respectively.  Rates of lensed GW in adLIGO were reassessed by \cite{ng18}, \cite{yang2019}.

 Under the geometrical optics approximation, the lensing effects  for  parameters estimation for individual  inspiral signal and the population of  black hole mergers  were addressed in \cite{cao14} and \cite{dai17} for ground-based  detectors, respectively.

\citet{Nakamura}
has first taken correct account of the diffraction effect in GW lensing and demonstrated that
when the lens is much lighter than $\sim 10^2\; M_{\odot}$, the diffraction is so effective that the geometric optics cannot be used anymore.
This subject has been advanced in the papers by \citet{Takahashi},\citet{Takahashi recent} and \cite{Liao2019}.

Given the  fact of laking  sky localization  ability of GW detectors,   previous  studies  of applications of  the lensed  gravitational waves   as a  astrophysical tool focus on the multi-messenger approach, namely people need to identify the host galaxy by their em counterpart to get the information of source  redshift and the lens potential.    The GW speed \cite{Fan17,Baker,Collett}, the Hubble constant \cite{Liao18} cosmic curvature \cite{li} and dark matter substructure \cite{liao}  have been studied with the information of  the time-delay from GW  and  redshift and location  from  EM signals  of a strong lens system.
The multi-messenger approach could only apply for the bns signal, while the major detection of GW should be BBH.

Recently, with the beat pattern of lensed GW alone, \cite{hou19b}  used matched filter strategy to determine the lens mass and the actual amplification of the point mass and SIS model  under the geometrical optics approximation  assuming known template for the lensed GWs.


In this paper we propose a  waveform independent methord to measure lens parameters and the actual amplification  by the ``beat" pattern of the strong lensed gravitational wave.
This paper is organized as follows: In section \ref{Lensing effect of chirp signal} we give the formulation of the lensing effect of GW, and introduce the characteristics of patterns in the lensed signals. In section \ref{measure} we provide the condition under which patterns can show up, and we discuss the relationship between the characters of patterns and parameters of both source and lens. Then we present a new approach to detect the information of gravitational lens and GW source. In section \ref{GW signal of CBC} we choose the CBC waveform and point-mass lens model as an example, to illuminate the characteristics of patterns we have discussed, and to show how our method works. And we give a summary in section \ref{discussion}.

\section{LENSING EFFECT OF GW}\label{Lensing effect of chirp signal}

In a gravitational lens system, we denote the distances between the source and lens, lens and observer, source and observer as $D_{LS}$, $D_L$ and $D_S$ respectively, 2 dimensional vector $\boldsymbol{\eta}$ denotes the distance of the source from the attachment of lens and observer, and $\boldsymbol{\xi}$ is a position vector on lens plane.

Following \citet{Takahashi}, for a monochromatic wave with frequency $f$, the effect of gravitational lens can be described by amplification factor $F(f)$ defined as:
\begin{equation}\label{}
F_{+,\times}(f)=\frac{\tilde{h}_{+,\times}^L(f)}{\tilde{h}_{+,\times}(f)}.
\end{equation}
The frequency  and polarization of the signal is not influenced by the lens and only the amplitude is remodulated.

The amplification factor is given by:
\begin{equation}\label{F(f)}
F(f,y)=\frac{D_S\xi_0^2}{D_LD_{LS}}\frac{f}{i}\int\mathrm{d}^2\boldsymbol{x}\exp[2\pi ift_d(\boldsymbol{x},\boldsymbol{y})],
\end{equation}
where $\boldsymbol{x}=\boldsymbol{\xi}/\xi_0, \boldsymbol{y}=\boldsymbol{\eta}D_L/\xi_0D_S$, and $\xi_0$ is an arbitrary normalization constant.

For the ground detectors, the detected signal  in time domain is then
\begin{equation}\label{}
h^L(t)=[A^+h^L_+(t)+A^{\times}h^L_\times(t)],
\end{equation}
where $A^+$ and $A^{\times}$ are antenna patten of the detector,
and $ h^L_{+,\times}(t)$  is the lensed  GW signal in time domain.
Hereafter we assume the antenna patten is a constant. 

\subsection{The beat pattern}\label{The effect of source and lens}
 When the wavelength of source signal is much larger than the Schwarzschild radius of the gravitational lens, there is no pattern but the amplification of the signal. While the wavelength is comparable with the Schwarzschild radius of lens, the patterns show up, and exist when wavelength is small enough to enter the realm of geometric optics. Here we  drive the exact condition for the patterns to show up for the strong lensed case.

To analyze the pattern on the lensed signal, we compute the envelope of the signal using Hilbert transform:
\begin{equation}\label{}
A^L(t)=\left|\frac{1}{\pi}\int\frac{h^L(\tau)}{t-\tau}\mathrm{d}\tau\right|,
\end{equation}
where $h^L(\tau)$  is the lensed GW signal.
 The position of peaks and valleys of the envelope are related to the lens mass $M_L$ and position $y$.
Note that for a typical lens (such as SIS and point mass model), the amplification factor is a function of the product of lens mass $M_L$ and the GW frequency $f$: $w=M_Lf$, so the lensed GW signal $\tilde{h}^L(f)$  in frequency domain can be written as:
\begin{equation}\label{lensh}
\tilde{h}^L(f)=F(w)\tilde{h}(f),
\end{equation}
So when the frequency changes slowly (e.g. $\dot{f} \ll f^2$ ), a necessary condition for peaks of the envelope in this case is:
\begin{equation}\label{}
\frac{\mathrm{d}F(w)}{\mathrm{d}w}\bigg|_{f=f_{p}}=0,
\end{equation}
where $f_p$ is the frequency of  peaks of the envelope.
So we have $f_{p}\propto1/M_L$ in this  case.
The  condition $\dot{f} \ll f^2$  is valid for all lensed CBCs in the wave region $\lambda \sim R_s$, and even valid for part of optical region, so we can  obtain the proportional coefficient in geometric optical region.

\begin{equation}\label{geo}
F(f)=\sum_j\mu_j^{1/2}e^{i2\pi ft_{dj}-i\pi n_j},
\end{equation}
meaning that discrete images interfere with each other, where $\mu_j$ is the magnification of the $j$th image, $t_{dj}=t_d(\boldsymbol{x_j},\boldsymbol{y})$, and $n_j=0, 1/2, 1$ when $\boldsymbol{x_j}$ is a minimum, saddle, and maximum point respectively of $t_d(\boldsymbol{x},\boldsymbol{y})$.

Suppose there are only two images, e.g., in the case of point-mass model and SIS model, if $t_0$ is the time of $n$th peak $f_p(n)$ and $t_0+\Delta t$ is the time of $(n+1)$th peak $f_p(n+1)$, they must satisfy,
\begin{equation}\label{peakcondition}
\int_{t_0}^{t_0+\Delta t}\left[f(\tau)-f(\tau-\Delta t_d)\right]\mathrm{d}\tau\propto1 ,
\end{equation}
where $\Delta t_d$ is the time delay of two lensed signals.
Expand Eq.~\ref{peakcondition} in Taylor series:
\begin{equation}\label{peakconditionTaylor}
\int_{t_0}^{t_0+\Delta t}\left[\dot f(\tau)\Delta t_d+\frac{1}{2}\ddot f(\tau)\Delta t_d^2+...\right]\mathrm{d}\tau\propto1.
\end{equation}
So in geometric region,
\begin{equation}\label{condition1}
\dot f(\tau)\Delta t_d\gg\frac{1}{2}\ddot f(\tau)\Delta t_d^2,
\end{equation}
Equation (\ref{peakcondition}) can be integrated as
\begin{equation}\label{delta f}
\left[f_p(n+1)-f_p(n)\right]\Delta t_d\propto1.
\end{equation}
So $\Delta f_{peak}\propto1/t_d\propto1/M_L$ when $\dot f(\tau)\Delta t_d\gg\frac{1}{2}\ddot f(\tau)\Delta t_d^2$.


\section{Measuring the Lens mass}\label{measure}
\subsection{The $f_{p}-n/M_L$ map}\label{map}
According to the geometric approximation  and \ref{delta f},
We obtain the relationship between $f_p$ and $M_L$, $n$, and $y$ when $\dot f(\tau)\Delta t_d\gg\frac{1}{2}\ddot f(\tau)\Delta t_d^2$:
\begin{equation}\label{final relation}
f_{p}=\alpha\frac{n}{\Delta t_d},
\end{equation}
Note that this equation is also valid to low frequency region as discussed in section \ref{The effect of source and lens}, just replace $t_d$ with its expression obtained in optical region. This equation implicates that the first peak appears when $f_{peak}=\alpha/\Delta t_d$, where $n=1$, and $\Delta t_d$ is given in the optical region. If condition (\ref{condition1}) is not satisfied, equation (\ref{peakcondition}) is always valid to describe the $f_{peak}-n/M_L$ map in high frequency region.



\subsection{The amplitudes ratio of beat and measurement of the lens mass}\label{am_ratio}
If we can measure appearing times or corresponding frequencies of more than two adjacent peaks, we can obtain $\Delta t_d$ from equation (\ref{final relation}) when condition (\ref{condition1}) is satisfied, or using (\ref{peakcondition}) when the condition is not satisfied. combining the timedelay information, we could measure the $M_L$ and $y$ by measuring the amplitudes ratio $r$ of valleys and peaks.

The unlensed CBC signal $h(t)$ could be written as:
\begin{equation}\label{waveform}
h(t)=\frac{a(M_c,\iota)}{D_L}b(t)e^{i(\Phi(t)+\phi_0)}.
\end{equation}
In geometric region, also suppose that there are only two images, the amplitude ratio of a valley at $t=t_1$ and its next adjacent peak at $t=t_2$ can be given by:
\begin{equation}\label{amplitude ration2}
r=\frac{\cos(\pi n_1)\sqrt{|\mu_1|}b(t_1)-\cos(\pi n_2)\sqrt{|\mu_2|}b(t_1-\Delta t_d)}{\cos(\pi n_1)\sqrt{|\mu_1|}b(t_2)+\cos(\pi n_2)\sqrt{|\mu_2|}b(t_2-\Delta t_d)},
\end{equation}
where $n_1$ and $n_2$ is defined in equation (\ref{geo}). This is because the peaks and valleys appear near the extreme of the two images where $(\Phi(t)+\phi_0)$ approximately equals to 0 or $\pi$, so the phase term is eliminated.

The form of $b(t)$ and $(\Phi(t)+\phi_0)$ can be determined by measuring the signal of the second image (after the coalescence of the first image), or by measuring the frequency of signal $f(t)$, since $f(t)$ is not influenced by the lens, as discussed in section \ref{Lensing effect of chirp signal}. From the information of $b(t)$, $(\Phi(t)+\phi_0)$ and the measurement of $r$, we can obtain $y$ from equation (\ref{amplitude ration2}), since $\mu$ is only a function of $y$. Then we can get $M_L$ from the expression of $\Delta t_d$, since $\Delta t_d$ is an expression of $M_L$ and $y$. Besides, we can also obtain the absolute magnification $\mu$, thus we can also measure the distance $D_L$ of the source.

\section{The application on inspiraling compact binary}\label{GW signal of CBC}


To demonstrate our approach,  we adopt the point mass lens model and  the newtonian approximation inspiral phase signal.
For the newtonian approximation inspiral phase signal, the  frequency of GW evolves as
\begin{equation}\label{}
f(t)=\frac{5}{4}\left(\frac{5GM_c}{c^3}\right)^{-5/8}(t_{coal}-t)^{-3/8}.
\end{equation}
In this case, condition (\ref{condition1}) $\dot f(\tau)\Delta t_d \gg \frac{1}{2}\ddot f(\tau)\Delta t_d^2$
 is equivalent to
\begin{equation}\label{condition2}
t_{coal}-t\gg\frac{11}{16}t_d.
\end{equation}

The amplification factor (\ref{F(f)})of the point mass lens model is analytically given by\cite{Takahashi}:
\begin{equation}\label{analytic}
\begin{split}
F(w,y)=&\exp\left\{\frac{\pi w}{4}+i\frac{w}{c}\left[\ln\left(\frac{w}{2}\right)-2\phi_m(y)\right]\right\}\\
&\times\Gamma\left(1-\frac{i}{2}w\right){}_1F_1\left(\frac{i}{2}w,1;\frac{i}{2}wy^2\right),
\end{split}
\end{equation}
where $w=8\pi M_Lf$, $\phi_m(y)=(x_m-y)^2/2-\ln x_m$ where $x_m=(y+\sqrt{y^2+4})/2$, and ${}_1F_1$ is the confluent hypergeometric function.
In geometric region, the approximation of amplification factor is:
\begin{equation}\label{Fhighf}
F(f)=\sqrt{|\mu^+|}-i\sqrt{|\mu^-|}e^{i2\pi f\Delta t_d},
\end{equation}
where
\begin{equation}\label{mu}
\mu^\pm=\frac{1}{2}\pm\frac{y^2+2}{2y\sqrt{y^2+4}},
\end{equation}
and
\begin{equation}\label{td}
\Delta t_d=4\frac{GM_L}{c^3}\left[\frac{y\sqrt{y^2+4}}{2}+\ln\left(\frac{\sqrt{y^2+4}+y}{\sqrt{y^2+4}-y}\right)\right].
\end{equation}

\begin{figure}[tbp]
\centering
\subfigure[3-3 $M_{\odot}$ binary black hole, $y=2.0$,, and $M_L=1000M_{\odot}$.]{
\label{} 
\includegraphics[width=8cm, angle=0]{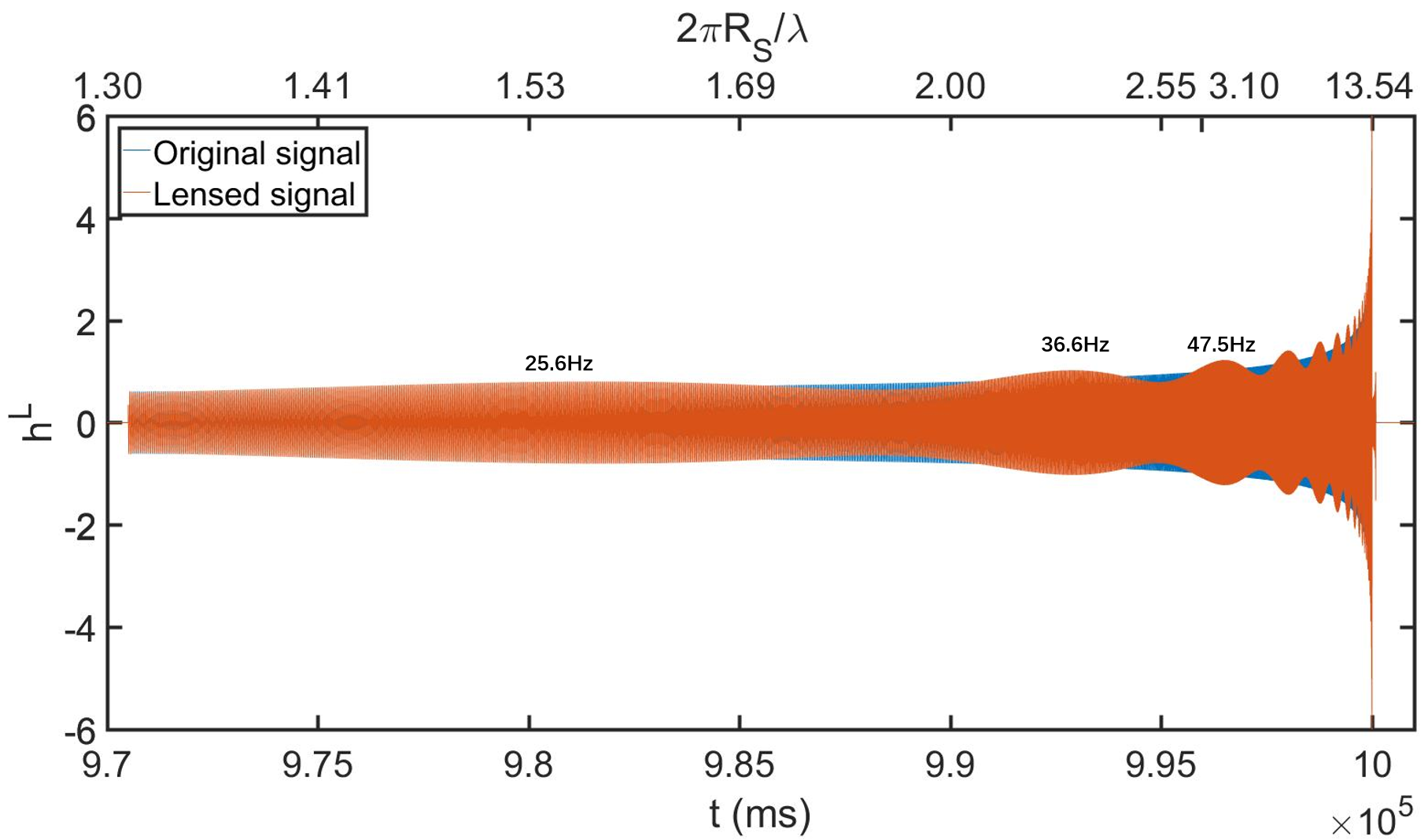}}
\subfigure[3-3 $M_{\odot}$ binary black hole, $y=2.0$,, and $M_L=1000M_{\odot}$.]{
\label{} 
\includegraphics[width=8cm, angle=0]{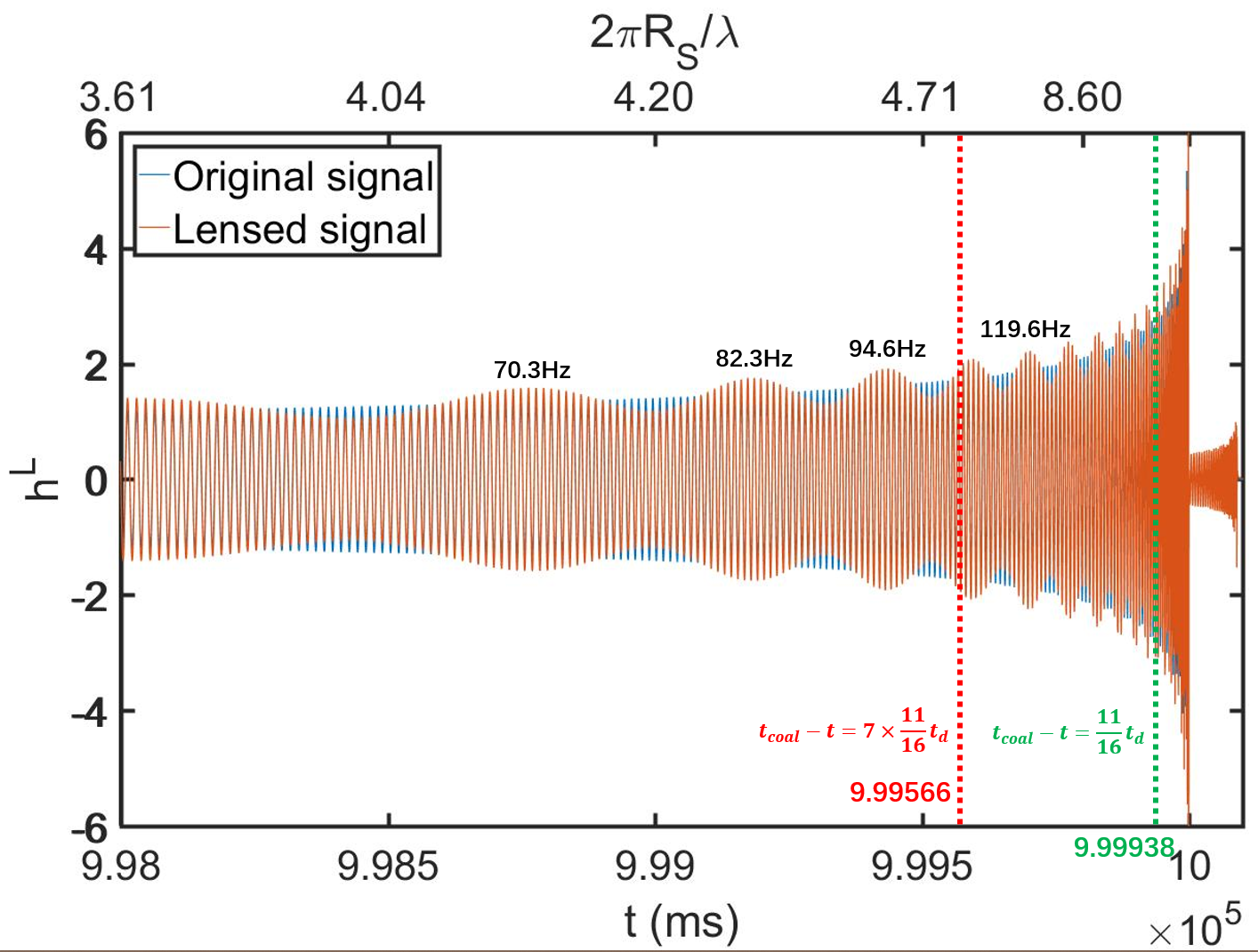}}
\hspace{1in} 
\subfigure[30-30 $M_{\odot}$ binary black hole, $y=3.0$, and $M_L=2000M_{\odot}$.]{
\label{} 
\includegraphics[width=8cm, angle=0]{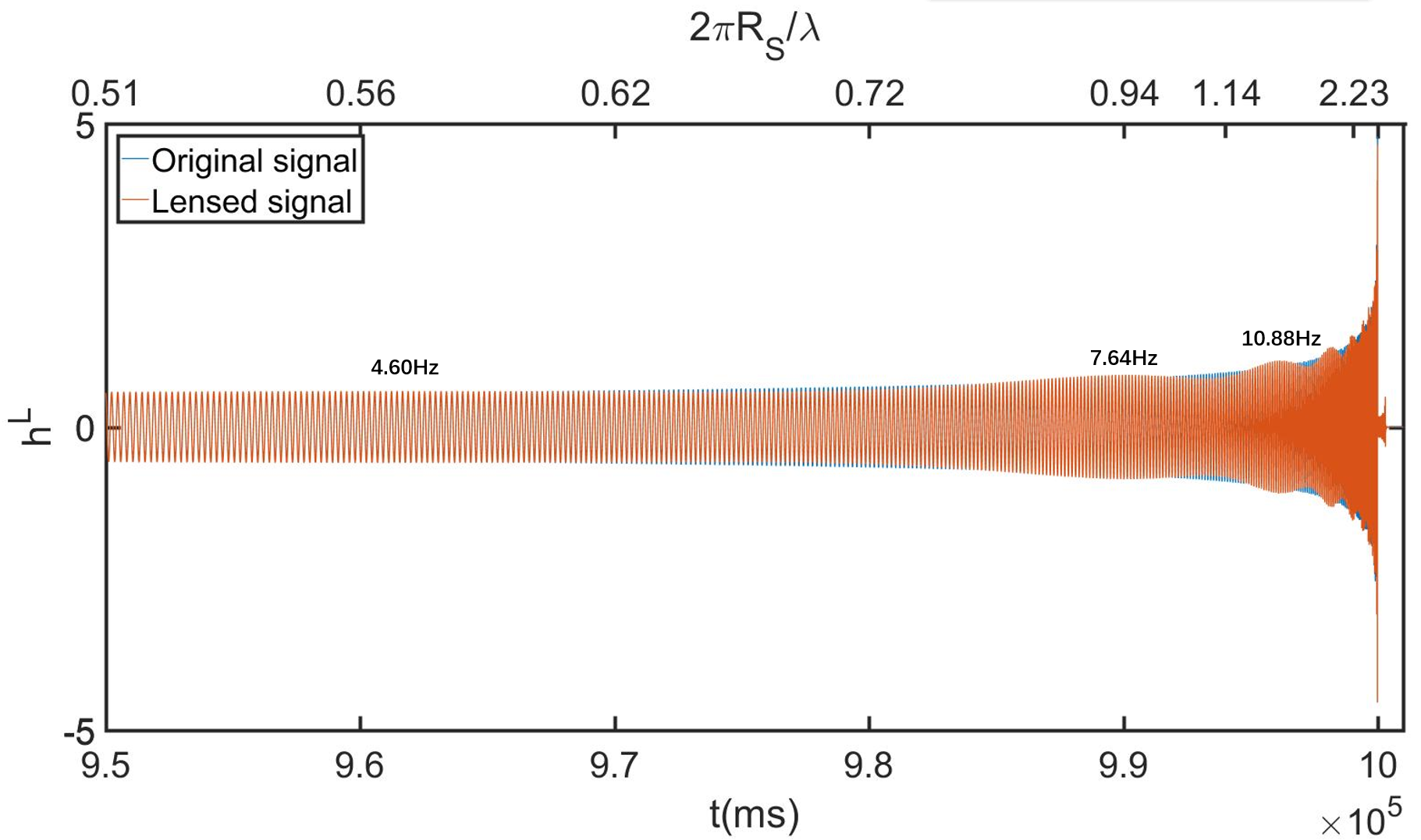}}
\subfigure[30-30 $M_{\odot}$ binary black hole, $y=3.0$, and $M_L=2000M_{\odot}$.]{
\label{} 
\includegraphics[width=8cm, angle=0]{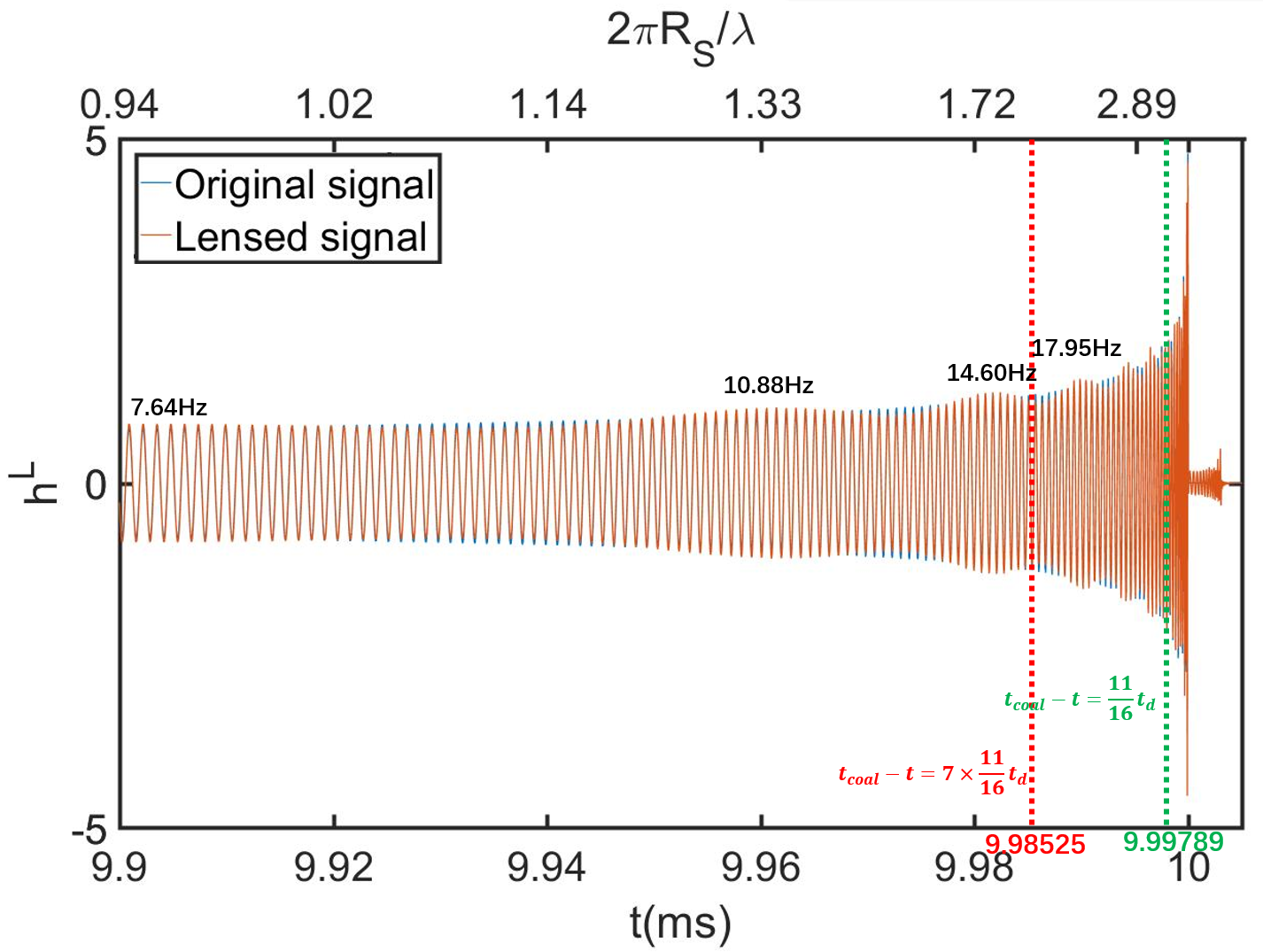}}
\caption{Examples of lensed waveform.}
\label{results} 
\end{figure}

Figure \ref{results} shows the ``beat" pattern for the 3-3 $M_{\odot}$ and 30-30 $M_{\odot}$ binary black holes. The point lens masses are 1000 $M_{\odot}$ and 2000 $M_{\odot}$, and $y$ are settled as 2 and 3, respectively. On the upper $x$-axis of these figures, we show $2\pi R_s/\lambda$ of the signal, which is the criterion of wave region and optical region, and the red lines delineate the boundaries for with condition (\ref{condition1}) is satisfied. From these results we can see that the patterns show up when the wavelength is comparable with the Schwarzschild radius of lens, and the positions of peaks before the red line is indicated by equation (\ref{final relation}).

\begin{figure}[tbp]
 \includegraphics[width=8cm, angle=0]{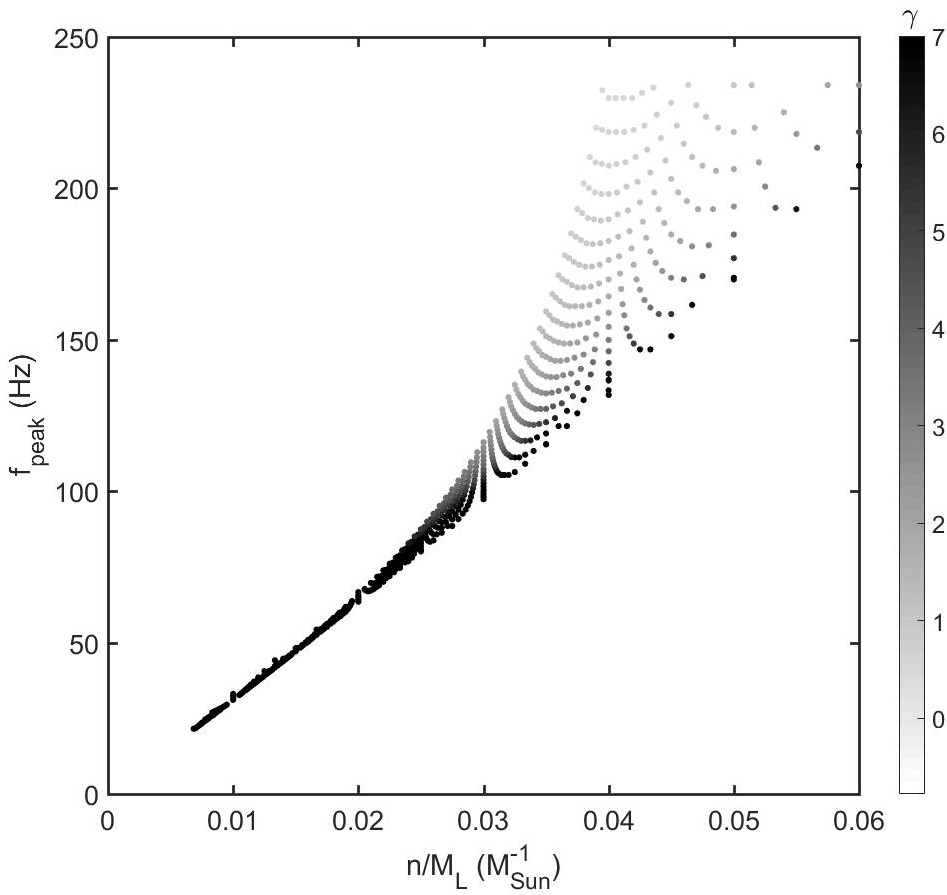}
 \caption{The relationship between $f_{p}$ and $n/M_L$, $y=5.0$, $M_L$ varies from $100M_\odot$ to $2000M_\odot$. Here $\gamma$ is defined as $t_{coal}-t=\gamma\frac{11}{16}t_d$.}
 \label{nM}
\end{figure}

To get the coefficients $\alpha$ of equation (\ref{final relation}) by fitting apprach, we change the mass of lens from 100 $M_{\odot}$ to 2000 $M_{\odot}$, and $y$ from 0.8 to 5.0.
As an example, we fix $y$ as 5.0, and plot the $f_{peak}-n/M_L$ map discussed in section \ref{map} in Figure \ref{nM}. We define $\gamma$ as $t_{coal}-t=\gamma\frac{11}{16}t_d$, and it is obvious that the larger the $\gamma$, the better the linear relationship discussed in equation (\ref{final relation}) is. And this figure also indicates that the proportional coefficients in high frequency region $\alpha$ are also valid in low frequency region, as discussed in section \ref{The effect of source and lens} and \ref{map}, which is $\alpha\approx1.12$ by fitting.

\begin{figure}[tbp]
 \includegraphics[width=8cm, angle=0]{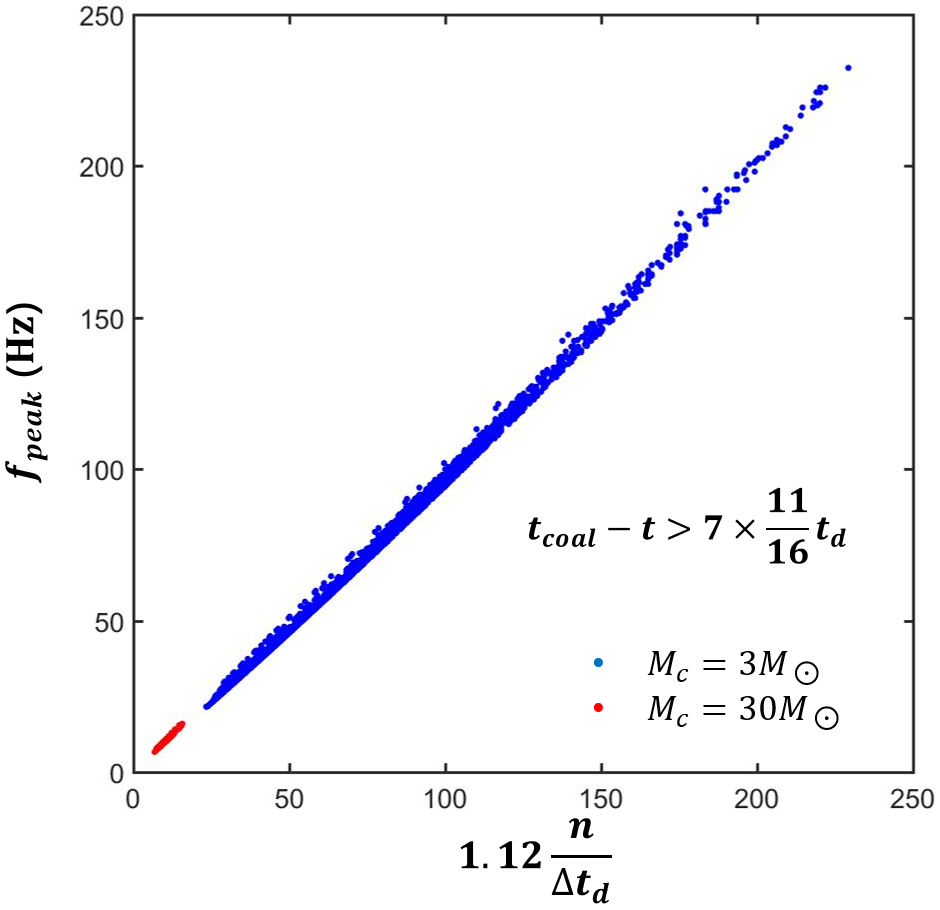}
 \caption{The verification of equation (\ref{final relation}) using $30-30M_\odot$ binary black holes. The lens mass also varies from $100M_\odot$ to $2000M_\odot$, and $y$ varies from 1.0 to 5.0.}
 \label{verification}
\end{figure}

We also test this formula with different sources and different $y$, e.g., $30-30M_\odot$ binary black holes, the lens mass also varies from $100M_\odot$ to $2000M_\odot$, and $y$ varies from 1.0 to 5.0. We also fit $\alpha$ by plotting $k-n/\Delta t_d$, and we find that $\alpha$ is also approximate to 1.12. The results are showed in Figure \ref{verification}.

\section{SUMMARY AND OUTLOOK}\label{discussion}
In this paper, we propose  a new method to measure the mass, position and the absolute magnification of gravitational lens for the strong lensed systerm with GW alone.
We focus on the `` betat" patterns of lensed GW, and we find the relationship between the frequency of the peaks $f_{p}$ of the envelope , the amplitude ratio of adjacent valleys and peaks $r$, and the mass $M_L$, position $y$ of gravitational lens: equations (\ref{peakcondition}), (\ref{final relation})
, and (\ref{amplitude ration2}),
where equation (\ref{peakcondition}) and (\ref{amplitude ration2}) is valid for optical region, and equation (\ref{final relation}) is valid for both wave and optical region as long as condition (\ref{condition1}) is satisfied. Thus by measuring the corresponding frequencies of more than two adjacent peaks, we can obtain $\Delta t_d$ from equation (\ref{final relation}) or (\ref{peakcondition}). Since the unlensed waveform with the normalized amplitude removed can be easily measured, i.e., $b(t)e^{i(\Phi(t)+\phi_0)}$ in waveform (\ref{waveform}), we can obtain the position of lens $y$ using equation (\ref{amplitude ration2}), then we can obtain the mass of lens $M_L$, the absolute magnification $\mu$, and therefore the distance of GW source $D_L$.

The point mass lens model and the BH-BH signals are used to demonstrate our methord. This systerm could be detectered in future ground based detectors.





\end{document}